\documentclass[epj]{svjour}
\usepackage{graphics,epsfig}
\newcommand{\beq}{\begin{equation}}
\newcommand{\eeq}[1]{\label{#1}\end{equation}}
\newcommand{\beqa}{\begin{eqnarray}}
\newcommand{\eeqa}[1]{\label{#1}\end{eqnarray}}
\newcommand{\eeqan}{\end{eqnarray}}
\newcommand{\CR}{\nonumber \\ }

\begin{document}
\title{Kaonic hydrogen versus the $K^{-}p$ low energy data}
\author{A.~Ciepl\'y\inst{1} \and J.~Smejkal\inst{2}
}                     
%
%
\institute{Nuclear Physics Institute, 250 68 \v{R}e\v{z}, Czech Republic 
\and Institute of Experimental and Applied Physics, 
Czech Technical University, Horsk\'{a} 3a/22, 128~00~Praha~2, 
Czech Republic}
\date{Received: date / Revised version: date}
%
\abstract{
We present an exact solution to the $K^{-}$-proton bound state problem 
formulated in the momentum space. The 1s level characteristics of 
the kaonic hydrogen are computed simultaneously with the available 
low energy $K^{-}p$ data. In the strong interaction sector 
the meson-baryon interactions are described by means of an effective 
(chirally motivated) separable potential and its parameters are fitted 
to the experimental data. 
} 
\maketitle
\section{Introduction}
\label{intro}
We developed a precise method of computing the meson-nuclear bound
states in momentum space. The method was already applied to
pionic atoms and its multichannel version was used to calculate the 1s level
characteristics of pionic hydrogen \cite{96CMa}. 
In the present work we aim at simultaneous 
description of both the 1s level kaonic bound state and the available 
experimental data for the $K^{-}p$ initiated processes. 

Until recently the old Deser-Trueman formula \cite{DTr} was used to determine 
the strong interaction energy shift and width (of the 1s level) in kaonic 
hydrogen from the $K^{-}p$ scattering length and vice versa. 
Recently, the Deser - Trueman relation was modified to include 
the isospin effects and electromagnetic corrections \cite{04MRR}. 
Our exact solution of the $K^{-}p$ bound state 
problem allows to check the precision and limitations of those approximate 
approaches. However, one should not forget that the strong interaction part 
of the scattering length is not a directly measured quantity and 
its determination from the scattering data is always model dependent. 

The treatment of the kaon-nucleon interaction at low energies requires a special 
care. Unlike the pion-nucleon interaction the $\bar{K}N$ dynamics is strongly 
influenced by the existence of the $\Lambda(1405)$ resonance, just below 
the $K^{-}p$ threshold. This means that the standard chiral perturbation 
theory is not applicable in this region. Fortunately, one can use 
non-perturbative coupled channel techniques to deal with the problem and 
generate the $\Lambda(1405)$ resonance dynamically. Such approach has proven 
quite useful and several authors have already applied it to various low 
energy meson-baryon processes \cite{95KSW}-\cite{06Oll}. 
Whether the recent experimental results on kaonic hydrogen from the  
DEAR collaboration \cite{05DEAR} are consistent with the older KEK results 
\cite{98KEK} and whether they fit into the picture drawn by the chiral 
models represents a question which is addressed by the theory \cite{06Oll,06BMN}
as well as by the coming SIDDHARTA experiment.

\section{Formalism}
\label{form}
Our approach to solving the meson-nuclear bound state problem in the presence 
of multiple coupled channels was given in Ref.~\cite{96CMa}. Here we just 
remark that the method is based on the construction of the Jost matrix 
and involves the solution of the Lippman-Schwinger equation for the transition 
amplitudes between various channels. Bound states in a specific 
channel then correspond to zeros of the determinant of the Jost matrix 
at (or close to) the positive part of the imaginary axis in the complex 
momentum plane. The zeros are computed iteratively and if only the point-like 
Coulomb potential is considered in the $K^{-}p$ channel the method reproduces 
the well known Bohr energy of the 1s level with a precision better 
than $0.1$~eV.

We follow the approach of Ref.~\cite{95KSW} when constructing the strong
interaction part of the potential matrix. In this model the $\Lambda(1405)$ 
resonance is generated dynamically by solving coupled Lippman-Schwinger 
equations with input effective (chirally motivated) potentials. The reader 
should note that our approach differs from the recently more popular 
on-shell scheme based on the Bethe-Salpeter equation, unitarity relation for the 
inverse of the $T$-matrix and on the dimensional regularization of the scalar 
loop integral \cite{01OMe}. Further, while the authors of Ref.~\cite{95KSW} 
restricted themselves only to the first six meson-baryon channels that are open 
at the $\bar{K}N$ threshold we employ all ten coupled meson-baryon channels: 
$K^-p$, $\bar{K}^{0}n$, $\pi^{0}\Lambda$, $\pi^{+}\Sigma^{-}$, $\pi^{0}\Sigma^{0}$, 
$\pi^{-}\Sigma^{+}$, $\eta \Lambda$, $\eta \Sigma^0$, $K^+\Xi^-$, and $K^0 \Xi^0$. 

The strong interaction potentials are constructed in such a way, that in 
the Born approximation they give the same (up to $\mathcal{O}(q^2)$) 
s-wave scattering lengths as are those derived from the underlying chiral 
lagrangian. Here we use them in the separable form 
\beqa
V_{ij}(k,k') & = & \sqrt{\frac{1}{2E_i}\frac{M_i}{\omega_i}}\:g_{i}(k)
\:\frac{C_{ij}}{f^2} \:g_{j}(k')\:\sqrt{\frac{1}{2E_j}\frac{M_j}{\omega_j}}~, 
\CR  g_{j}(k) & = & \frac{1}{1+(k/ \alpha_{j})^2}~,
\eeqa{poten}
in which the momenta $k$ and $k'$ refer to the meson-baryon c.m. system 
in the $i$ and $j$ channels, respectively, and the kinematical factors 
$\sqrt{M_j/(2 E_j \omega_j)}$ guarantee a proper relativistic flux 
normalization with $E_j$, $M_j$ and $\omega_j$ denoting the meson energy 
and the baryon mass and energy in the c.m. system of channel $j$. 
The off shell form factors $g_{j}(k)$ introduce the inverse range 
radii $\alpha_{j}$ that characterize the radius of interactions in 
various channels. Finally, the parameter $f$ stands for the pseudoscalar 
meson decay constant in the chiral limit and the coupling matrix $C_{ij}$ 
is determined by chiral SU(3) symmetry and includes terms up to the second 
order in the meson c.m.~kinetic energies. For the first six channels the 
couplings $C_{ij}$ were listed in \cite{95KSW} and we intend to publish 
the remaining coefficients in a more elaborate paper \cite{08CSm}. For 
illustration, we show just the coupling of the elastic $K^{-}p$ process, 
\beqa
C_{K^{-}p,K^{-}p} & = & -E_K - \frac{E_{K}^{2}-m_{K}^{2}}{2M_0} 
             +(F^2+{D^2 \over 3})\frac{E_{K}^{2}}{2M_0} + \CR
           & + & 4 m^2_K (b_D+b_0) - E^2_K (2 d_D + 2d_0+d_1)~.
\eeqa{C11}
Here $m_K$ and $E_K$ denote the kaon mass and energy in the center-of-mass 
frame, $M_0$ stands for the baryon mass in the chiral limit, and the parameters 
$F$, $D$, and $b$'s and $d$'s represent coupling constants that appear  
in the underlying chiral lagrangian (see \cite{95KSW} for more details).
The origin and relevance of the various terms present in Eq.~(\ref{C11}) 
was discussed thoroughly in Ref.~\cite{05BNW}. In general, the coefficients 
$C_{ij}$ include contributions from the meson-baryon contact interactions 
as well as the direct and crossed Born terms. However, in contrast to 
\cite{05BNW} our model is based on the static (heavy) nucleon approximation 
adopted by the authors of Ref.~\cite{95KSW} in which the underlying 
lagrangian is expressed in a fixed reference frame.

The potential of Eq.~(\ref{poten}) is used not only when solving the bound 
state problem but we also implement it in the standard Lippman-Schwinger 
equation and compute the low energy $\bar{K}N$ cross sections and branching 
ratios from the resulting transition amplitudes.

\section{$\bar{K}N$ data fits}
\label{fits}
The parameters of the chiral lagrangian which enter the coefficients 
$C_{ij}$ and the inverse range radii $\alpha_{j}$ determining the off-shell 
behavior of the potentials are to be fitted to the experimental data. Before 
performing the fits we reduce the number of the fitted parameters 
in the following way. First, the axial couplings $D$ and $F$ were already 
fixed in the analysis of semileptonic hyperon decays \cite{99Rat}, 
$D = 0.80$, $F = 0.46$ ($g_{A} = F + D = 1.26$). 
Then, we fix the couplings $b_D$ and $b_F$ to satisfy the approximate 
Gell-Mann formulas for the baryon mass splittings, 
\beqa
M_{\Xi} - M_N            &=& - 8 b_F (m_K^2 -m_\pi^2)~, \CR
M_{\Sigma} - M_{\Lambda} &=& \frac{16}{3} b_D (m_K^2 -m_\pi^2)~, 
\eeqan
which gives $b_D = 0.064$ GeV$^{-1}$ and $b_F = -0.209$ GeV$^{-1}$. 
Similarly, we determine the coupling $b_0$ and the baryon chiral mass 
$M_0$ from the relations for the pion-nucleon sigma term $\sigma_{\pi N}$ 
and the proton mass,
\beqa
\sigma_{\pi N} &=& -2 m_\pi^2(2b_0+b_D+b_F)  ~, \CR
M_p            &=& M_0-4m_K^2( b_0+ b_D-b_F)-2 m_\pi^2 (b_0+2b_F) ~.
\eeqa{sigma}
Since the value of the pion-nucleon $\sigma$-term is not well determined we 
enforce four different options, $\sigma_{\pi N} = 20\:-\:50$ MeV, which 
cover the interval of the values considered by various authors.
Finally, we reduce the number of the inverse ranges $\alpha_{j}$ 
to only five: $\alpha_{KN}$, $\alpha_{\pi \Lambda}$, 
$\alpha_{\pi \Sigma}$, $\alpha_{\eta \Lambda /\Sigma}$ (for both the 
$\eta \Lambda$ and $\eta \Sigma$ channels), and $\alpha_{K \Xi}$. 
This leaves us with 11 free parameters: the five 
inverse ranges, the meson-baryon chiral coupling $f$, and five more 
low energy constants from the second order chiral lagrangian denoted by 
$d_D$, $d_F$, $d_0$, $d_1$, and $d_2$.

The fitted low energy $\bar{K}N$ data include the three precisely measured 
threshold branching ratios \cite{81Mar}
\beqa
\gamma&=&\frac{\sigma(K^-p\rightarrow \pi^+\Sigma^-)}
{\sigma(K^-p\rightarrow \pi^-\Sigma^+)}=2.36\pm 0.04, \CR
R_c&=&\frac{\sigma(K^-p\rightarrow \hbox{charged particles})}
{\sigma(K^-p\rightarrow \hbox{all})}=0.664\pm0.011, \CR
R_n&=&\frac{\sigma(K^-p\rightarrow \pi^0\Lambda)}
{\sigma(K^-p\rightarrow \hbox{all neutral states})}=0.189\pm 0.015,
\eeqa{rates}
and $K^- p$-initiated total cross sections. For the later ones we consider 
only the experimental data taken at the kaon laboratory momenta 
$p_{LAB} = 110$ MeV (for the $K^- p$, $\bar{K^0}n$, $\pi^{+} \Sigma^{-}$,
$\pi^{-} \Sigma^{+}$ final states) and at $p_{LAB} = 200$ MeV (for 
the same four channels plus $\pi^0 \Lambda$ and $\pi^{0} \Sigma^{0}$). 
Our results show that the inclusion of the cross section data taken 
at other kaon momenta is not necessary since the fit 
at just $1-2$ points fixes the cross section 
magnitude and the energy dependence is reproduced nicely by the model. 
Finally, we include the DEAR results \cite{05DEAR} on the strong interaction 
shift $\Delta E_N$ and the width $\Gamma$ of the 1s level in kaonic hydrogen:
\beq
\Delta E_{N}(1s)= (193 \pm 43)  \hbox{ eV,}\;\;\;
\Gamma(1s) = (249 \pm 150) \hbox{ eV~.}
\eeq{Katom}  
Thus, we end up with a total of 15 data points in our fits.

\begin{table*}[t]
\caption{The fitted $\bar{K}N$ threshold data}
\begin{center}
\begin{tabular}{ccccccc}
$\sigma_{\pi N}$ [MeV] & $\chi^{2}/N$ & $\Delta E_{N}$ [eV] & $\Gamma$ [eV]& $\gamma$ & $R_c$ & $R_n$ \\ \hline
 20           & 1.33  & 232     & 725      & 2.366   & 0.657 & 0.191 \\
 30           & 1.36  & 262     & 697      & 2.365   & 0.657 & 0.190 \\
 40           & 1.37  & 253     & 710      & 2.370   & 0.657 & 0.189 \\
 50           & 1.40  & 266     & 708      & 2.370   & 0.658 & 0.190 \\ \hline
 exp          &   -   & 193(43) & 249(150) & 2.36(4) & 0.664(11) & 0.189(15)
\end{tabular}
\end{center}
\label{fits}
\end{table*}

\begin{table*}
\caption{Chiral lagrangian parameters ($b_0$ and $d$'s in $1/$GeV):}
\begin{center}
\begin{tabular}{cccccccccc}
$\sigma_{\pi N}$ [MeV] & $b_0$ & $M_0$ [MeV] & $a_{\pi N}^{+}$ [$m_{\pi}^{-1}$]& $f$ [MeV] & $d_0$ & $d_D$ & $d_F$ & $d_1$ & $d_2$ \\ \hline
 20                    & -0.190 & 997 & -0.016 & 108.6 & -0.385 & -0.368 & -0.817 & 0.396 & 0.152  \\
 30                    & -0.321 & 864 &  0.001 & 100.0 & -0.354 & -0.206 & -0.522 & 0.406 & -0.211  \\
 40                    & -0.453 & 729 &  0.006 & 108.9 & -0.484 & -0.151 & -0.459 & 0.448 & -0.280  \\
 50                    & -0.584 & 594 &  0.007 & 108.8 & -0.747 & -0.092 & -0.429 & 0.567 & -0.349  \\ \hline
 27~\cite{95KSW}       & -0.279 & 910 & -0.002 &  94.5 & -0.40  & -0.24  & -0.43  & 0.28  & -0.62   
\end{tabular}
\end{center}
\label{param}
\end{table*}

Our results are summarized in Tables \ref{fits}-\ref{alphas}. The first table 
shows the results of our $\chi^{2}$ fits compared with the relevant experimental 
data. The resulting $\chi^{2}$ per data point indicate satisfactory 
fits. It is worth noting that their quality and the computed values 
do not depend much on the exact value of the $\sigma_{\pi N}$ term. 
Tables \ref{param} and \ref{alphas} show the fitted parameters 
of the chiral lagrangian and the inverse range parameters $\alpha_j$. 
The last rows in the tables compare our values with those determined 
in Ref.~\cite{95KSW}. We remind the reader that the parameter $b_0$ and the 
baryon mass in the chiral limit were not fitted to the data and are given 
in the second and third column of Table \ref{param} only to visualize their 
respective values corresponding to the selected $\sigma_{\pi N}$ term. 
The $\pi N$ isospin-even scattering 
length $a_{\pi N}^{+}$ shown in the fourth column of Table \ref{param} was 
not included in our fits either but we feel that its presentation 
is important and deserves some comments. 

The goal of the present work was to check the compatibility of the DEAR 
kaonic hydrogen data with the low energy $K^{-}p$ cross sections and branching 
ratios. Therefore, we have not included in our fits the $\Lambda$(1405) 
mass spectrum and other processes considered e.g.~in Ref.~\cite{06Oll}. 
In fact, the low energy constants involved in the fits should be also 
constrained by other observables calculated within the framework 
of ChPT involving the same meson-baryon lagrangian. The spectrum of baryon 
masses and the $\pi N$ isospin-even scattering length may come to one's mind 
in this respect. The later quantity to order $q^3$ is given by \cite{93BKM}:
\beqa
  a^+_{\pi N} & = & {{1} \over {4 \pi (1+m_\pi /M_N)}} \times \CR
  & \times & \biggl[ {{m^2_\pi} \over {f^2}}(-2b_D - 2b_F - 4b_0 + d_D + d_F + 2 d_0)-
  \CR & - &  {{m^2_\pi} \over {f^2}}{{g^2_A} \over {4 M_N}}
            +  {{3 g^2_A m^3_\pi} \over {64 \pi f^4}}  \biggr]~.
\eeqa{a_piN}
Since the experimental value of $a^+_{\pi N}$ is practically consistent with 
zero, $a_{0+}^+=-(0.25\pm 0.49) \cdot 10^{-2}\,m_\pi^{-1}$ \cite{99Sch}, 
it is encouraging to note the mostly negative signs of the $d$-parameters 
that cancel the positive contributions due to the $b$ terms and the $q^{3}$
correction represented by the last term in Eq.~(\ref{a_piN}). 
As a smaller $\sigma_{\pi N}$ term means a smaller absolute 
value of the negative parameter $b_0$ (and hence a smaller positive 
contribution due to the $b_0$ term in $a_{\pi N}^{+}$) it is not surprising 
that the computed $\pi N$ scattering length is becoming negative 
for too low $\sigma_{\pi N}$ terms. Anyway, it is interesting that our fits 
aimed at the $\bar{K}N$ interactions allow for so good reproduction 
of the $\pi N$ quantity. Specifically, the parameter set obtained in the fit 
for $\sigma_{\pi N}=30$ MeV gives $a_{\pi N}^{+}$ in a nice agreement with  
experiment while the $\chi^{2}/N$ is only slightly inferior to our best fit. 
Many other authors (e.g. \cite{95KSW} or \cite{06Oll}) include 
the $a_{\pi N}^{+}$ value directly in their fits. The $d$ couplings 
(of the second order chiral lagrangian) 
contribute to the contact meson-baryon interactions in the second order 
of meson momenta. Although our fits confirm their mostly negative signs 
it is difficult to come to any conclusions concerning their values. 
The fact that even the sign of $d_2$ is not well determined in our analysis 
speaks for itself. 

We have also tried to perform fits with the $b$ parameters taken from the
analysis of the baryon mass spectrum \cite{97BMe} and with only the current 
algebra (Weinberg-Tomozawa) term contributing to the $C_{ij}$ coefficients 
(the approach adopted in Ref.~\cite{98ORa}). Unfortunately, we were not 
able to achieve satisfactory results in those cases.
Thus, we conclude that the low 
energy constants derived in the analysis of baryon masses are not suitable 
in the sector of meson-baryon interactions and that the inclusion of 
the $q^{2}$ terms is necessary for a good description of the $\bar{K}N$ data. 
The later point is in agreement with the analysis of Ref.~\cite{05BNW}.

\begin{table}
\caption{Inverse range parameters $\alpha_j$ (in MeV):}
\begin{center}
\begin{tabular}{cccccc}
$\sigma_{\pi N}$ [MeV] & $\alpha_{KN}$ & $\alpha_{\pi \Lambda}$ & 
  $\alpha_{\pi \Sigma}$ & $\alpha_{\eta \Lambda /\Sigma}$ & 
  $\alpha_{K \Xi}$ \\ \hline
  20    & 610 & 209 & 570 & 1100 & 530 \\ 
  30    & 647 & 262 & 535 & 308 & 21 \\ 
  40    & 653 & 320 & 618 & 281 & 89 \\ 
  50    & 594 & 370 & 610 & 342 & 124 \\ \hline
 27~\cite{95KSW} & 760 & 300 & 450 &  -  &  -  
\end{tabular}
\end{center}
\label{alphas}
\end{table}

The inverse range parameters given in Table \ref{alphas} are in line with 
our expectations. The values corresponding to the open channels $\bar{K}N$, 
$\pi \Lambda$ and $\pi \Sigma$ seem to be well determined and show only 
a moderate dependence on the adopted value of the $\sigma_{\pi N}$ term. 
In general, the ranges obtained for the open channels correspond to the 
t-channel exchanges that are believed to dominate the interactions.
On the other hand the range of interactions in the closed channels is not
well defined in the fits and the fitted values $\alpha_{\eta \Lambda /\Sigma}$  
and $\alpha_{K \Xi}$ exhibit relatively large statistical errors. This 
feature also justifies our use of only one range parameter for both 
$\eta$ channels. 

In Figure 1 we present the low energy $K^{-}p$ initiated cross sections 
calculated using our best fit with $\sigma_{\pi N} = 20$ MeV. The results 
obtained for the other adopted values of $\sigma_{\pi N}$ are quite similar, 
therefor we decided to not include them in the figure. Though we declined 
from using all experimental data in our fits and took only the data points 
available for the selected kaon laboratory momenta $p_{LAB} = 110$ MeV 
and $p_{LAB} = 200$ MeV, the description of the data is quite good. 
Specifically, we do not observe the lowering of the calculated cross sections 
in the elastic $K^{-}p$ channel reported by Borasoy et al.~\cite{05BNW} 
for their fits including the kaonic hydrogen characteristics. Though our 
$K^{-}p$ cross sections are also slightly below the experimental data 
the difference is not significant. In addition, the inclusion of 
electromagnetic corrections discussed in Ref.~\cite{05BNW} should partly 
improve the description for the lowest kaon momenta.

\begin{figure*}
\includegraphics[width=\textwidth]{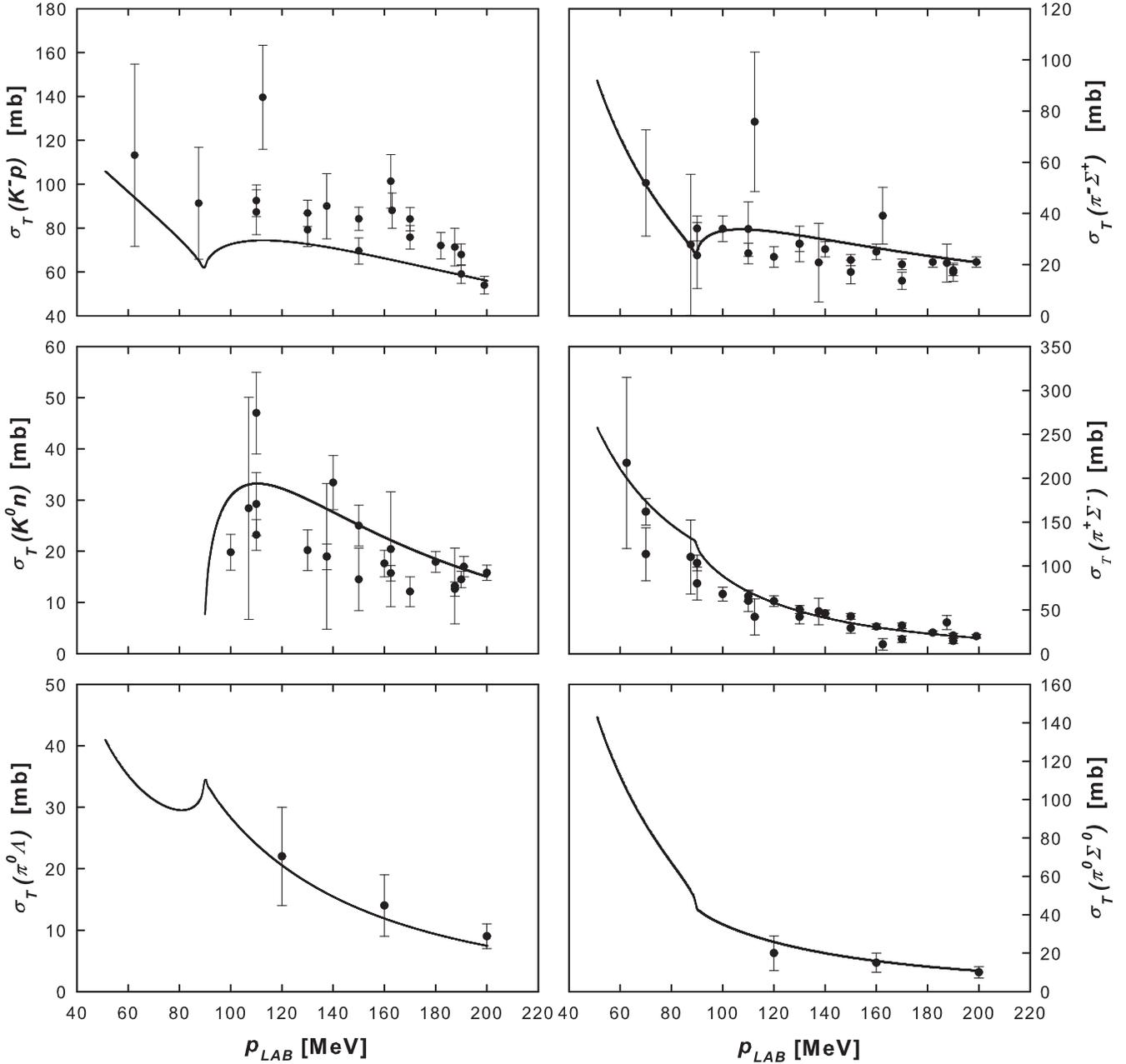} 
\caption{Total cross sections for $K^{-}p$ scattering and reactions to the 
meson-baryon channels open at low kaon laboratory momenta $p_{LAB}$. 
The experimental data are the same as those compiled 
in Fig.~1 of Ref.~\cite{95KSW}.}
\label{fig:1}
\end{figure*}

Finally, let us turn our attention to the calculated characteristics 
of the 1s level in kaonic hydrogen. The strong interaction 
energy shift of the 1s level in kaonic hydrogen is reproduced well but we 
were not able to get a satisfactory fit of the 1s level energy width as our 
results are significantly larger than the experimental value. This result is 
in line with the conclusions reached by Borasoy, Meissner and Nissler 
\cite{06BMN} on the basis of their comprehensive analysis of the $K^{-}p$ 
scattering length from scattering experiments. However, when considering 
the interval of three standard deviations and also the older KEK 
results \cite{98KEK} (which give less precise but larger width) we cannot 
conclude that kaonic hydrogen measurements contradict the other low energy 
$\bar{K}N$ data. 

In Table \ref{DT} we compare our results (for $\sigma_{\pi N}=20$ 
and $50$ MeV) for the 1s level characteristics in kaonic hydrogen 
with the approximate values determined from the $K^{-}p$ scattering 
lengths $a_{K^{-}p}$. The later quantity is obtained from the multiple 
channel calculation that uses the same parametrization of the strong 
interaction potential, Eq.~(\ref{poten}). The 1s level complex energies are 
shown for: the standard Deser-Trueman formula (DT) \cite{DTr}, the modified 
Desert-Trueman formula (MDT) \cite{04MRR} (see also Ref.~\cite{05BNW} 
for the relations used to obtain the DT and MDT values) and our ``exact" 
solution of the bound state problem. The results obtained for $\sigma_{\pi N}=30$ 
and $40$ MeV are quite similar (with almost identical $K^{-}p$ scattering 
lengths), so we did not include them in the table in order to keep 
the presentation more transparent. Since the numerical precision 
of determining the bound state energy by our method is better than $0.1$~eV, 
the discrepancy between the ``MDT" and the ``exact" values should 
be attributed to higher order corrections not considered in the 
derivation of MDT. In fact, the correction due to Coulomb interaction is taken 
only in its leading order in the MDT formula. The inclusion of more terms 
of the relevant geometric series would bring the MDT value into a better 
agreement with our exact solution \cite{07Rus}.

\begin{table}
\caption{Precision of the Deser-Trueman formula. The complex energies 
$\Delta E_{N} - ({\rm i}/2)\Gamma$ are given for the approximate 
DT and MDT formulas and compared with our computed ``exact" values.}
\begin{center}
\begin{tabular}{c|cc}
                   
$a_{K^{-}p}$ [fm] &  & $\Delta E_{N} - ({\rm i}/2)\Gamma$ [eV] \\ \hline \hline
                         &  DT   & $207 - ({\rm i}/2)\,832$ \\
$-0.50 + {\rm i}\: 1.01$ &  MDT  & $251 - ({\rm i}/2)\,714$ \\
                         & exact & $232 - ({\rm i}/2)\,725$ \\ \hline
                         &  DT   & $247 - ({\rm i}/2)\,830$ \\   
$-0.60 + {\rm i}\: 1.01$ &  MDT  & $285 - ({\rm i}/2)\,689$ \\
                         & exact & $266 - ({\rm i}/2)\,708$
\end {tabular}
\end{center}
\label{DT}
\end{table}

\section{Conclusions}
\label{end}
We have computed the characteristics of kaonic hydrogen 
exactly and compared the results (the 1s level energy shift and width) 
with the values determined from the $K^{-}p$ scattering length by means 
of using the standard Deser-Trueman formula and its modified version 
which includes the corrections due to electromagnetic effects. 
It looks that the approximate DT formula gives the 1s energy level 
strong interaction shift and width about $10\%$ and $15\%$ off the exactly 
computed values, respectively. Although the modified DT formula does much 
better job on account of the width the energy level shift remains about 
$10\%$ off the exact value that lies approximately in the middle between 
the DT and MDT values. In view of the current level of the experimental 
precision the use of the modified DT formula is sufficient. Nevertheless, 
the situation may change after the coming SIDDHARTA experiment 
being prepared in Frascati.

An effective chirally motivated separable potential was used 
in simultaneous fits of the low energy $K^{-}p$ cross sections, 
the threshold branching ratios and the characteristics of kaonic hydrogen. 
The fits are quite satisfactory except the 1s level energy width being much 
larger than the experimental value. In view of the fact that the experimental 
precision of the kaonic hydrogen data is still rather low one cannot say 
that the data contradict the chirally motivated model used to describe 
the low energy meson-baryon interactions. However, as the opposite statement 
cannot be made either we should wait for the new experiment to clarify 
the situation.

\vspace*{4mm}
{\bf Acknowledgement:} A.~C. acknowledges the financial support from the GA~AVCR 
grant A100480617. The work of J.~S. was supported by the Research Program 
{\it Fundamental experiments in the physics of the microworld} No.~6840770029 
of the Ministry of Education, Youth and Sports of the Czech Republic. 
  
%
%

\end{document}